\documentclass[twocolumn,preprintnumbers,showpacs,pra,a4paper,superscriptaddress]{revtex4}

\usepackage{graphicx}
\usepackage{color}
\usepackage{amsmath}
\usepackage{amssymb}
\usepackage{subfigure}
\usepackage{epsfig}
\usepackage[normalem]{ulem}

\begin{document}
\title{Modelling the effect of charge noise on the exchange interaction between spins}

\author{M. J. Testolin}\affiliation{Centre for Quantum Computer Technology, School of Physics, The University of Melbourne, Victoria 3010, Australia.}
\author{J. H. Cole}\affiliation{Centre for Quantum Computer Technology, School of Physics, The University of Melbourne, Victoria 3010, Australia.}
\affiliation{Institut f\"ur Theoretische Festk\"orperphysik und DFG-Center for Functional Nanostructures (CFN), 
Universit\"at Karlsruhe, 76128 Karlsruhe, Germany.}
\author{L. C. L. Hollenberg}\affiliation{Centre for Quantum Computer Technology, School of Physics, The University of Melbourne, Victoria 3010, Australia.}

\begin{abstract}
We describe how the effect of charge noise on a pair of spins coupled via the exchange interaction can be calculated by modelling charge fluctuations as a random telegraph noise process using probability density functions.  We develop analytic expressions for the time dependent superoperator of a pair of spins as a function of fluctuation amplitude and rate.  We show that the theory can be extended to include multiple fluctuators, in particular, spectral distributions of fluctuators.  These superoperators can be included in time dependent analyses of the state of spin systems designed for spintronics or quantum information processing to determine the decohering effects of exchange fluctuations.
\end{abstract}

\pacs{74.40.+k,71.70.Gm,05.40.-a}

\maketitle

\section*{Introduction}
The exchange interaction is of increasing importance in the study of controllable quantum mechanics using solid-state systems.  As well as being fundamentally important in manybody physics, it is this interaction which is often used to mediate spin flips or entanglement in spintronics and quantum information processing (QIP)~\cite{Kane_Nature_393_133_1998, Loss_PRA_57_120_1998, Vrijen_PRA_62_012306_2000, Wolf_Science_294_1488_2001, Friesen_PRB_67_121301_2003, Zutic_RevModPhys_76_323_2004, Hollenberg_Phys_Rev_B_74_045311_2006}.  For these reasons there has been considerable study recently on the origin and control of the exchange interaction~\cite{deSousa_PRA_64_042307_2001, Koiller_PRL_88_027903_2002, Wellard_PRB_72_085202_2005, Coish_PhysRevB_75_161302_2007,Hill_PhysRevLett_98_180501_2007, Testolin_PhysRevA_76_012302_2007}.  For applications involving the time varying control of the exchange interaction, such as QIP, the stability in time of this interaction is of crucial importance.  As the origin of the exchange interaction is essentially the overlap of electron wave functions, the interaction strength is sensitive to the local charge environment.  Recent work~\cite{Hu_PRL_96_100501_2006} has shown that the dependence of the exchange interaction is approximately linearly dependent on fluctuations in the local electric field.

In this paper we develop a general framework with which the effect of these fluctuations can be analytically included in time dependent calculations of the state of a spin system.  Using the superoperator formalism~\cite{Kubo_1957,Fano_1957,Zwanzig_1960,Fano_1963,Zwanzig_1964,Ernst90}, we derive an expression for a pair of spins as a function of time, depending on the exchange fluctuation amplitude and rate.  The extension to multiple fluctuators, in particular, spectral distributions of fluctuators is also considered.  The formalism can be used to investigate the effect of exchange fluctuations on spintronics, quantum control schemes and specifically quantum error correction (QEC).

\section{The noise model}
\label{sect:the_model}
We begin by studying the exchange coupling Hamiltonian in the presence of a single charge fluctuator with the aim of understanding the decohering effects of the fluctuator.  The Hamiltonian for the process is
\begin{equation}
H\left(t\right)=J\left(t\right)\mbox{\boldmath$\sigma$\unboldmath}_1\cdot\mbox{\boldmath$\sigma$\unboldmath}_2.
\label{eqn:exchange_hamiltonian}
\end{equation}
The exchange coupling $J\left(t\right)$ varies in time due to a random telegraph noise (RTN) process, $\eta\left(t\right)$ and we assume a net effect of the form 
\begin{equation}
J\left(t\right)=J_0+\alpha\eta\left(t\right),
\label{eqn:exchange_coupling}
\end{equation}
where  $\eta\left(t\right)$ describes the fluctuator.  This RTN process couples with strength $\alpha$ (ultimately dependent on the distance between the coupled spins and the fluctuator) to the bare exchange term, $J_0$.  The time evolution of the system can then be described by the density matrix master equation
\begin{equation}
\dot{\rho}\left(t\right)=-i\left[H\left(t\right),\rho\left(t\right)\right],
\label{eqn:density_matrix_evolution}
\end{equation}
where $\rho\left(t\right)$ is the density matrix of the system.  Additional terms can be added to this master equation to also model non-unitary evolution, such as decohering processes.

As a matter of convenience we may re-express the system evolution in superoperator form.  In superoperator form the density matrix is given a vector representation, denoted by $\vec{\rho}\left(t\right)$, by transforming the matrix into a single column, one row at a time~\cite{Barnett90}.  A superoperator $\mathbf{P}\left(t\right)$ contains all the evolution of the system (both unitary and non-unitary)
\begin{equation}
\dot{\vec{\rho}}\left(t\right)=\mathbf{P}\left(t\right)\vec{\rho}\left(t\right).
\label{eqn:superoperator_form}
\end{equation}
For purely Hamiltonian evolution the superoperator $\mathbf{P}\left(t\right)$ can be written down in terms of $H\left(t\right)$ and the identity operator
\begin{equation}
\mathbf{P}\left(t\right)= -i\left[H\left(t\right)\otimes I-I\otimes H\left(t\right)^T\right].
\label{eqn:Form_of_superoperator_P}
\end{equation}
The superoperator simplifies to
\begin{equation}
\mathbf{P}\left(t\right)= -iJ\left(t\right)\mbox{\boldmath$\sigma$\unboldmath}_{\rm H},
\label{eqn:Form_of_our_superoperator_P}
\end{equation}
for the Hamiltonian we consider.  Here, $\mbox{\boldmath$\sigma$\unboldmath}_{\rm H}$ is the Heisenberg interaction in superoperator form.  If the Hamiltonian is time independent, then the superoperator {\bf P} is also time independent and the density matrix at some time $t$ is
\begin{eqnarray}
\vec{\rho}\left(t\right)&=&e^{\mathbf{P}t}\vec{\rho}\left(t_0\right),\\
&\equiv&\mathbf{q}\left(t\right)\vec{\rho}\left(t_0\right),
\label{eqn:rho_at_time_t}
\end{eqnarray}
given an initial state $\vec{\rho}\left(t_0\right)$.  We show how this time independent formalism is relevant to our problem shortly.

The RTN process, $\eta\left(t\right)$, is modelled as in Ref.~\cite{Mottonen_PRA_73_022332_2006}.  The noise fluctuates randomly between -1 and 1 with the frequency of the fluctuations controlled by the correlation time $1/\lambda$.  Here, $\lambda$ is the typical frequency of jump times, where the jump time instants are,
\begin{equation}
t_i=\sum_{j=1}^i-\frac{1}{\lambda}\ln\left(p_j\right),
\label{eqn:jump_time_instants}
\end{equation}
and the $p_j$ are random numbers such that $p_j\in\left(0,1\right)$.  The noise process $\eta\left(t\right)$ is described as
\begin{equation}
\eta\left(t\right)=\left(-1\right)^{\sum_i\Theta\left(t-t_i\right)}\eta\left(0\right),
\label{eqn:RTN_model}
\end{equation}
where $\Theta\left(t\right)$ is the Heaviside step function, and $\eta\left(t\right)$ can fluctuate between $\pm \eta\left(0\right)$.  We choose $|\eta\left(0\right)|=1$ and control the coupling strength via $\alpha$ as in Eq.~\ref{eqn:exchange_coupling}.

The density matrix evolution for our system can be found by numerically averaging over many such noise histories $\eta\left(t\right)$ to obtain the correct system dynamics.  For an initial state $\rho\left(t_0\right)$,
\begin{equation}
\rho\left(t\right)=\lim_{N\to\infty}\frac{1}{N}\sum_{k=1}^N U_k\rho\left(t_0\right)U_k^\dagger,
\label{eqn:rho_in_terms_of_U}
\end{equation}
where the $\left\{U_k\right\}$ are the evolution operators for trajectories $\eta_k\left(t\right)$.  Since the Hamiltonian (Eq.~\ref{eqn:exchange_hamiltonian}) commutes with itself at all times,
\begin{equation}
\left[H\left(t_0\right),H\left(t\right)\right]=0,
\label{eqn:commutation_relation}
\end{equation}
the $\left\{U_k\right\}$ may be expressed as
\begin{equation}
U_k\left(t,t_0\right)=U^{-}_{k}\left(t_-\right)U^{+}_{k}\left(t_+\right),
\label{eqn:evolution operator_for given_noise_history}
\end{equation}
where $t_-$ and $t_+$ describe the total time the fluctuator exists in the -1 and +1 states respectively for a particular noise history, and
\begin{equation}
U_{k}^{\pm}\left(t\right)=e^{-i\left(J_0\pm\alpha\right)\mbox{\boldmath$\sigma$\unboldmath}_1\cdot\mbox{\boldmath$\sigma$\unboldmath}_2t}.
\label{eqn:evolution_operator_for particular_value_of_noise}
\end{equation}
Using the result of Eq.~\ref{eqn:rho_at_time_t} for the superoperator form of a density matrix governed by a time independent Hamiltonian we re-express Eq.~\ref{eqn:rho_in_terms_of_U} such that
\begin{equation}
\vec{\rho}\left(t\right)=\lim_{N\to\infty}\frac{1}{N}\sum_{k=1}^N \mathbf{q}_{k}^{-}\left(t_-\right)\mathbf{q}_{k}^{+}\left(t_+\right)\vec{\rho}\left(t_0\right).
\label{eqn:rho_in_superoperator_form}
\end{equation}
The ensemble averaged superoperator, $\mathbf{Q}\left(t\right)$, is the average of all the individual trajectory superoperators $\mathbf{q}_{k}\left(t\right)$,
\begin{equation}
\mathbf{Q}\left(t\right)=\lim_{N\to\infty}\frac{1}{N}\sum_{k=1}^N \mathbf{q}_{k}^{-}\left(t_-\right)\mathbf{q}_{k}^{+}\left(t_+\right).
\label{eqn:big_Q}
\end{equation}
This implies that $\mathbf{Q}\left(t\right)$ may be constructed by numerically averaging over many noise histories.  The averaging is crucial in obtaining the correct system dynamics, as the RTN is a stochastic process and so there are many unique noise trajectories.  Averaging over these noise trajectories results in non-unitary evolution despite the Hamiltonian being strictly unitary.

Conversely, it is possible to derive $\mathbf{Q}\left(t\right)$ analytically by describing the stochastic RTN using an appropriate probability density function (PDF).  By considering all unique $\mathbf{q}_{k}\left(t\right)$ as a function of the average fluctuator state $\xi=|\eta\left(0\right)|\left(t_+-t_-\right)/T$, weighted by a PDF giving the occurrence likelihood of the average fluctuator state, and integrating this over all possible $\xi$, the resulting expression for $\mathbf{Q}\left(t\right)$ is
\begin{equation}
\mathbf{Q}\left(t\right)=\int_\xi\mathbf{q}_{\xi}\left(t\right)\Omega\left(\xi,T\right)\,d\xi.
\label{eqn:integral_over_average_fluctuator_state}
\end{equation}
Here, $\mathbf{q}_{\xi}\left(t\right)$ is the unique individual superoperator corresponding to a particular value of $\xi$ and $\Omega\left(\xi,T\right)$ is the PDF, which determines the probability that during the time interval $T$, the average fluctuator state is $\xi$.  In section~\ref{sect:PDF} we show how to specify the PDF, so that we can use it to analytically determine $\mathbf{Q}\left(t\right)$ in section~\ref{sect:Determining_Q}.

\section{Calculating the probability density function}
\label{sect:PDF}
The statistical properties of an RTN process have been studied extensively in the context of reliability theory, alternating renewal processes and queueing theory~\cite{Takacs_ActaMathHung_8_169_1957,Nelson:95,Barlow:96,Aven:99,Perry_QueueingSyst_33_369_1999,Zacks_JApplProbab_41_497_2004}.  In our case, we are specifically interested in the probability of the RTN spending a certain fraction of the observation period in a particular state.  The PDF for an RTN signal fluctuating between the states 0 and +1 is given by~\cite{Zacks_JApplProbab_41_497_2004} as
\begin{equation}
p\left(\tau,T\right)=\lambda e^{-\lambda T}\sqrt{\frac{\tau}{T-\tau}}I_1\left[2\lambda\sqrt{\tau\left(T-\tau\right)}\right],
\label{egn:pdf_derived from Zacks_paper}
\end{equation}
where $I_1$ is the modified Bessel function of the first kind.  This PDF assumes the initial state is +1, and that \emph{at least} a single fluctuation occurs.  Here, $\tau$ is used to describe the time spent in the state 0 and $T$ is the duration of the process we are considering.  The parameter $\lambda$ characterises the fluctuator rate as before.  Properly normalised the PDF is,
\begin{equation}
p\left(\tau,T\right)=\frac{\lambda}{2}\sqrt{\frac{\tau}{T-\tau}}\frac{I_1\left[2\lambda\sqrt{\tau\left(T-\tau\right)}\right]}{\sinh^2\left(\frac{\lambda T}{2}\right)}.
\label{egn:pdf_normalised}
\end{equation}
We could equally describe a process which begins in the state 0, with $T-\tau$ describing the time spent in this state.  Assuming at least a single fluctuation occurs, the full PDF is obtained by averaging over both possible starting states
\begin{equation}
p^\prime\left(\tau,T\right)=\frac{1}{2}\left[p\left(\tau,T\right)+p\left(T-\tau,T\right)\right].
\label{eqn:pdf_averaging}
\end{equation}
We may re-express this PDF in terms of the mean fluctuator state $\xi$, where $\xi\in\left[-1,1\right]$.  Taking care to preserve the normalisation, the PDF for an RTN process of duration $T$ assuming at least one fluctuation occurs is
\begin{eqnarray}
\Omega_{>0}\left(\xi,T\right)&=&\frac{T}{2}p^\prime\left[\frac{T}{2}\left(\xi+1\right),T\right],\\
&=&\frac{\lambda T}{4}\frac{I_1\left(\lambda T\sqrt{1-\xi^2}\right)}{\sqrt{1-\xi^2}\sinh^2\left(\frac{\lambda T}{2}\right)}.
\label{eqn:pdf_general}
\end{eqnarray}

The case where no fluctuations occur must be treated separately.  In this case we expect $\xi$ to be either of $\pm 1$.  The properly normalised PDF for this case can be described using two delta functions,
\begin{equation}
\Omega_0\left(\xi,T\right)=\frac{1}{2}\left[\delta\left(\xi-1\right)+\delta\left(\xi+1\right)\right].
\label{eqn:pdf_no_fluctuations}
\end{equation}
The full, general PDF is constructed by appropriately weighting $\Omega_0\left(\xi,T\right)$ and $\Omega_{>0}\left(\xi,T\right)$, with the fluctuation probability given by the Poisson distribution
\begin{equation}
p_k\left(\lambda T\right)=\frac{e^{-\lambda T}\left(\lambda T\right)^k}{k!},
\label{eqn:Poisson_distribution}
\end{equation}
where $k$ denotes the number of fluctuations, such that
\begin{equation}
\Omega\left(\xi,T\right)=p_0\left(\lambda T\right)\Omega_0\left(\xi,T\right)+p_{>0}\left(\lambda T\right)\Omega_{>0}\left(\xi,T\right),
\label{eqn:full_pdf_general_construction}
\end{equation}
and $p_{>0}\left(\lambda T\right)=1-p_0\left(\lambda T\right)$.  After simplification, the resulting PDF is
\begin{multline}
\Omega\left(\xi,T\right)=\frac{e^{-\lambda T}}{2}\left[\delta\left(\xi-1\right)+\delta\left(\xi+1\right)\right]\\
+\frac{\lambda T}{e^{\lambda T}-1}\frac{I_1\left(\lambda T\sqrt{1-\xi^2}\right)}{\sqrt{1-\xi^2}}.
\label{eqn:full_pdf_general}
\end{multline}

In what follows we examine the three limiting cases of the PDF and use these to construct an approximate PDF.  The approximate PDF provides greater physical insight when working within these limits.

Examining the two limiting cases of the PDF $\Omega_{>0}\left(\xi,T\right)$, the fast and slow fluctuator limits and combining them with $\Omega_0\left(\xi,T\right)$, leads to a simplified expression which approximates $\Omega\left(\xi,T\right)$.  We begin by considering the slow fluctuator limit $\lambda\rightarrow 0$ for the distribution describing at least one fluctuation, $\Omega_{>0}\left(\xi,T\right)$.  This is the regime where no more than one fluctuation occurs.  In this limit
\begin{equation}
I_a\left(x\right)\sim\frac{1}{\Gamma\left(a+1\right)}\left(\frac{x}{2}\right)^a,
\label{eqn:I_asymptotic_form_lambda_small}
\end{equation}
and
\begin{equation}
\sinh\left(x\right)=x+\mathcal{O}\left(x^3\right).
\label{eqn:sinh_approximation_lambda_small}
\end{equation}
This reduces the PDF to
\begin{eqnarray}
\Omega_{>0}\left(\xi,T\right)&\approx&\frac{1}{2},\\
&\equiv&\tilde{\Omega}_1\left(\xi,T\right).
\label{eqn:pdf_lambda_0}
\end{eqnarray}
This uniform distribution implies that a fluctuation is just as likely to occur at any time during the system evolution.

The limit $\lambda\rightarrow\infty$ represents a fast fluctuator.  In this regime
\begin{equation}
I_a\left(x\right)\sim\frac{1}{\sqrt{2\pi x}}e^x,
\label{eqn:I_asymptotic_form_lambda_large}
\end{equation}
and
\begin{equation}
\sinh\left(x\right)\approx\frac{e^x}{2},
\label{eqn:sinh_approximation_lambda_large}
\end{equation}
which reduces the PDF to
\begin{eqnarray}
\Omega_{>0}\left(\xi,T\right)&\approx&\sqrt{\frac{\lambda T}{2\pi}}e^{-\frac{\lambda T\xi^2}{2}}\left(1+\frac{3}{4}\xi^2\right),\\
&\approx&\sqrt{\frac{\lambda T}{2\pi}}e^{-\frac{\lambda T\xi^2}{2}}+\mathcal{O}\left(\xi^2\right).
\label{eqn:pdf_lambda_infinity}
\end{eqnarray}
In this limit $\xi$ will be small, therefore making the substitution $\mu=1/\sqrt{\lambda T}$, we find the PDF to be Gaussian about the origin,
\begin{eqnarray}
\Omega_{>0}\left(\xi,T\right)&\approx&\frac{1}{\mu\sqrt{2\pi}}e^{-\frac{\xi^2}{2\mu^2}},\\
&\equiv&\tilde{\Omega}_{>1}\left(\xi,T\right).
\label{eqn:pdf_lambda_infinity_Gaussian}
\end{eqnarray}
which we expect intuitively.  We note that this approach is similar to that used by Happer and Tam when considering the Gaussian limit of rapid spin exchange in alkali vapors~\cite{Happer_PRA_16_1877_1977}.

Weighting these two limiting cases and the PDF describing no fluctuations using the Poisson distribution as before, allows us to construct an approximate PDF
\begin{multline}
\Omega\left(\xi,T\right)\approx p_0\left(\lambda T\right)\Omega_0\left(\xi,T\right)+p_1\left(\lambda T\right)\tilde{\Omega}_1\left(\xi,T\right)\\
+p_{>1}\left(\lambda T\right)\tilde{\Omega}_{>1}\left(\xi,T\right),
\label{eqn:approx_full_pdf_general_construction}
\end{multline}
where $p_{>1}\left(\lambda T\right)=1-p_0\left(\lambda T\right)-p_1\left(\lambda T\right)$.  This approximate $\Omega\left(\xi,T\right)$ provides nice analytic solutions for $\mathbf{Q}\left(t\right)$ in each of the three interesting fluctuator regimes.  While this is only an approximation to the exact solution (Eq.~\ref{eqn:full_pdf_general}), it can provide more physical insight, as will become apparent later.

\section{Using the PDF to determine $\mathbf{Q}\left(t\right)$}
\label{sect:Determining_Q}
The superoperator $\mathbf{Q}\left(t\right)$ can be derived analytically via Eq.~\ref{eqn:integral_over_average_fluctuator_state} using the PDFs determined in the previous section.  Of particular interest is the non-unitary part of the superoperator.

The non-unitary superoperator can be found by expanding the superoperator into a unitary and non-unitary part, such that $\mathbf{Q}\left(t\right)=\mathbf{Q}^{\left(\rm u\right)}\left(t\right)\mathbf{Q}^{\left(\rm nu\right)}\left(t\right)$.  The evolution in the absence of a fluctuator is contained within the unitary part,
\begin{equation}
\mathbf{Q}^{\left(\rm u\right)}\left(t\right)=e^{-iJ_0\mbox{\boldmath$\sigma$\unboldmath}_{\rm H}t}, 
\label{eqn:unitary_superoperator_evolution}
\end{equation}
whilst the effect of the charge fluctuator is contained within the non-unitary part $\mathbf{Q}^{\left(\rm nu\right)}\left(t\right)$.  Note that these two parts can be factored out due to the commutation relation (Eq.~\ref{eqn:commutation_relation}).  We now determine the non-unitary parts of the superoperator for $\Omega\left(\xi,T\right)$ and its various approximations.

Beginning with the case where no fluctuations occur and the PDF is given by $\Omega_0\left(\xi,T\right)$, as in Eq.~\ref{eqn:pdf_no_fluctuations}, we find
\begin{equation}
\mathbf{Q}^{\left(\rm nu\right)}_0\left(t\right)=\cos\left(\alpha\mbox{\boldmath$\sigma$\unboldmath}_{\rm H}t\right),
\label{eqn:Qnu_no_fluctuations}
\end{equation}
where $\mbox{\boldmath$\sigma$\unboldmath}_{\rm H}$ is the Heisenberg superoperator introduced earlier.  When there is at least one fluctuation (see Eq.~\ref{eqn:pdf_general}) the resulting form of the superoperator is
\begin{equation}
\mathbf{Q}^{\left(\rm nu\right)}_{>0}\left(t\right)=\frac{\cos\left[\sqrt{\left(\alpha\mbox{\boldmath$\sigma$\unboldmath}_{\rm H}t\right)^2-\left(\lambda T\right)^2}\right]-\cos\left(\alpha\mbox{\boldmath$\sigma$\unboldmath}_{\rm H}t\right)}{2\sinh^2\left(\frac{\lambda T}{2}\right)}.
\label{eqn:Qnu_general}
\end{equation}
Examining the limiting cases of the general PDF we find that for the slow fluctuator (see Eq.~\ref{eqn:pdf_lambda_0})
\begin{equation}
\mathbf{\tilde{Q}}^{\left(\rm nu\right)}_1\left(t\right)=\frac{\sin\left(\alpha\mbox{\boldmath$\sigma$\unboldmath}_{\rm H}t\right)}{\alpha\mbox{\boldmath$\sigma$\unboldmath}_{\rm H}t},
\label{eqn:Qnu_lambda_0_limit}
\end{equation}
and in the fast fluctuator limit (see Eq.~\ref{eqn:pdf_lambda_infinity_Gaussian})
\begin{equation}
\mathbf{\tilde{Q}}^{\left(\rm nu\right)}_{>1}\left(t\right)=e^{-\left(\alpha\mu\mbox{\boldmath$\sigma$\unboldmath}_{\rm H}t\right)^2/2}.
\label{eqn:Qnu_lambda_infinite_limit}
\end{equation}
It should be noted that this final superoperator corresponds exactly to that which would be obtained using the Lindbladian formalism~\cite{Lindblad_Commun_Math_Phys_48_119_1976,Gardiner:91} if a decoherence operator of the form $L=\frac{\alpha}{\sqrt{\lambda}}\mbox{\boldmath$\sigma$\unboldmath}_1\cdot\mbox{\boldmath$\sigma$\unboldmath}_2$ was included.  From this one can deduce that the fast fluctuator limit is equivalent to purely Markovian decoherence due to interaction with the environment via an exchange like two-qubit decoherence channel.  This is in contrast to conventional dephasing which is modelled using two independent $\sigma_Z$ channels, one for each qubit.  This distinction is particularly important as it implies that exchange fluctuations due to environmental charge fluctuations introduce \emph{correlated} errors which can have important implications for fault-tolerant QEC~\cite{Aliferis_QIC_6_97_2006}.

Using the previous results, we can determine $\mathbf{Q}^{\left(\rm nu\right)}\left(t\right)$ for the full weighted PDF's in both the approximate and exact cases.  The exact PDF given in Eq.~\ref{eqn:full_pdf_general} yields
\begin{multline}
\mathbf{Q}^{\left(\rm nu\right)}\left(t\right)=e^{-\lambda T}\cos\left(\alpha\mbox{\boldmath$\sigma$\unboldmath}_{\rm H}t\right)+\frac{2}{e^{\lambda T}-1}\\
\times\left\{\cos\left[\sqrt{\left(\alpha\mbox{\boldmath$\sigma$\unboldmath}_{\rm H}t\right)^2-\left(\lambda T\right)^2}\right]-\cos\left(\alpha\mbox{\boldmath$\sigma$\unboldmath}_{\rm H}t\right)\right\},
\label{eqn:Qnu_full_exact}
\end{multline}
whilst for the approximate PDF given in Eq.~\ref{eqn:approx_full_pdf_general_construction} we find
\begin{multline}
\mathbf{Q}^{\left(\rm nu\right)}\left(t\right)\approx e^{-\lambda T}\cos\left(\alpha\mbox{\boldmath$\sigma$\unboldmath}_{\rm H}t\right)+\lambda Te^{-\lambda T}\frac{\sin\left(\alpha\mbox{\boldmath$\sigma$\unboldmath}_{\rm H}t\right)}{\alpha\mbox{\boldmath$\sigma$\unboldmath}_{\rm H}t}\\
+\left(1-e^{-\lambda T}-\lambda Te^{-\lambda T}\right)e^{-\left(\alpha\mu\mbox{\boldmath$\sigma$\unboldmath}_{\rm H}t\right)^2/2}.
\label{eqn:Qnu_approximate}
\end{multline}
In general it is difficult to graphically compare these analytic forms of the superoperator to the numerical result.  However, it is possible in this case, as the superoperator $\mathbf{Q}^{\left(\rm nu\right)}\left(t\right)$ is a sparse matrix with the same underlying structure of the $\mbox{\boldmath$\sigma$\unboldmath}_{\rm H}$ superoperator which defines it.  It follows from the definition of $\mbox{\boldmath$\sigma$\unboldmath}_{\rm H}$ (see Eq.~\ref{eqn:Form_of_our_superoperator_P}) that the only non-zero matrix elements of the Heisenberg superoperator are $\pm 2$.  Consequently, a comparison of the resulting non-zero matrix element of $\mathbf{Q}^{\left(\rm nu\right)}\left(t\right)$, denoted $Q_{\rm NU}$, proves effective in determining the agreement between the analytic (exact and approximate) superoperators and exact numerical solution for the superoperator.
\begin{figure}[h]
\centerline{\includegraphics{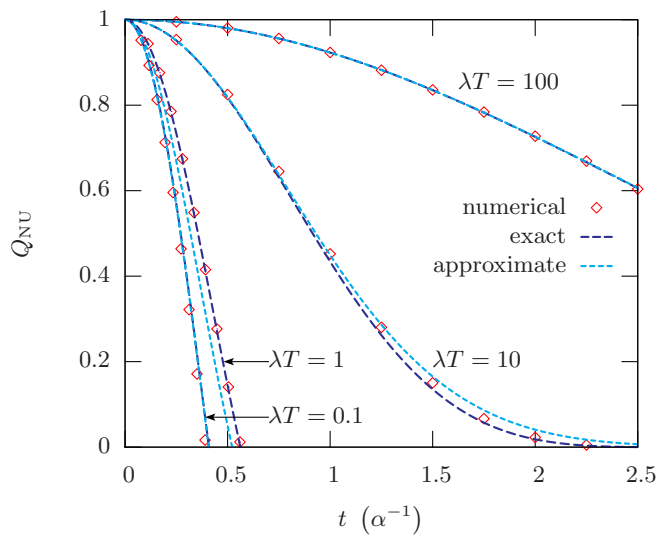}}
\caption{Comparison of the analytical [exact (Eq.~\ref{eqn:Qnu_full_exact}) and approximate (Eq.~\ref{eqn:Qnu_approximate})] and exact numerical (simulated from Eq.~\ref{eqn:big_Q}) solutions of $\mathbf{Q}^{\left(\rm nu\right)}\left(t\right)$.  Plotted is the non-zero matrix element of $\mathbf{Q}^{\left(\rm nu\right)}\left(t\right)$, denoted $Q_{\rm NU}$, as a function of time for a range of fluctuator rates, which span each of the three limiting regimes.  The results show very good agreement between all three solutions, with the analytic approximation deviating only slightly when the fluctuator rate is on the time scale of $T$.}
\label{fig:Qnu_analytic_vs_numeric}
\end{figure}

The results, as a function of time for a range of fluctuator rates, shown in Fig.~\ref{fig:Qnu_analytic_vs_numeric}, reveal very good agreement between the exact analytic and numerical results for all rates $\lambda$.  The approximate solution also matches closely, particularly in the slow and fast fluctuator limits.  Slight deviations from the exact solution can be seen when the fluctuations occur on the time scale of the process we are considering $\left(\lambda T\approx 1\right)$.  In this regime the contribution from the uniform distribution $\tilde{\Omega}_1\left(\xi,T\right)$ is at its maximum and approximately on par with contributions from the other two distributions.  The deviation from the exact results do not come as a surprise, as the approximate PDF is constructed from contributions due to 0, 1 or many fluctuations.  Adding contributions from 2, 3 and more fluctuations would reduce this discrepancy.

We now present the generalisation of the single fluctuator formalism to multiple fluctuators in the following section.

\section{Multiple fluctuators}
\label{sect:multiple_fluctuators}
Extending this formalism to multiple fluctuators is straight forward and provides a method for the treatment of many physically realistic scenarios.  The total ensemble averaged superoperator, $\mbox{\boldmath$\Lambda$\unboldmath}\left(t\right)$, for $N$ fluctuators is just the product of all the individual ensemble averaged superoperators, $\mathbf{Q}\left(t\right)$, such that
\begin{equation}
\mbox{\boldmath$\Lambda$\unboldmath}\left(t\right)=\mathbf{Q}^{\left(\rm u\right)}\left(t\right)\prod_{i=1}^N \mathbf{Q}^{\left(\rm nu\right)}_i\left(t\right).
\label{eqn:multiple_fluctuators}
\end{equation}
This result is useful for a finite number of fluctuators each with known strength and rate.  However, in most instances only the spectral distribution in strength and rate will be known and therefore Eq.~\ref{eqn:multiple_fluctuators} offers no further insight.  By considering all possible unique superoperators $\mathbf{Q}^{\left(\rm nu\right)}\left(\alpha_i,\lambda_i,t\right)$ weighted by their probability of occurrence $p_i$ (where $p_i\in\left[0,1\right]$), in a similar way to the method used to construct $\mathbf{Q}\left(t\right)$ in Eq.~\ref{eqn:integral_over_average_fluctuator_state}, we may re-express Eq.~\ref{eqn:multiple_fluctuators} as
\begin{equation}
\mbox{\boldmath$\Lambda$\unboldmath}\left(t\right)=\mathbf{Q}^{\left(\rm u\right)}\left(t\right)\prod_{i=1}^M \left[\mathbf{Q}^{\left(\rm nu\right)}\left(\alpha_i,\lambda_i,t\right)\right]^{Np_i},
\label{eqn:multiple_fluctuators_weighted}
\end{equation}
where in general there are $M$ possible fluctuator types and $N$ fluctuators.  We would like to interpret the $p_i$ as a spectral distribution function in $\alpha_i$ and $\lambda_i$.  By expressing $\mbox{\boldmath$\Lambda$\unboldmath}$ as a sum of logarithms
\begin{equation}
\mbox{\boldmath$\Lambda$\unboldmath}\left(t\right)=\mathbf{Q}^{\left(\rm u\right)}\left(t\right)\exp\left\{N\sum_{i=1}^M p_i\ln\left[\mathbf{Q}^{\left(\rm nu\right)}\left(\alpha_i,\lambda_i,t\right)\right]\right\},
\label{eqn:multiple_fluctuators_sum_log}
\end{equation}
and extending the definition of $\mbox{\boldmath$\Lambda$\unboldmath}$ to the continuum, we may replace the $p_i$ with a spectral distribution function $S\left(\alpha,\lambda\right)$ such that
\begin{multline}
\mbox{\boldmath$\Lambda$\unboldmath}\left(t\right)=\mathbf{Q}^{\left(\rm u\right)}\left(t\right)\\
\times\exp\left\{N\iint S\left(\alpha,\lambda\right)\ln\left[\mathbf{Q}^{\left(\rm nu\right)}\left(\alpha,\lambda,t\right)\right]\,d\alpha\,d\lambda\right\},
\label{eqn:multiple_fluctuators_integral_log}
\end{multline}
ensuring that the spectral distribution function is properly normalised
\begin{equation}
\iint S\left(\alpha,\lambda\right)\,d\alpha\,d\lambda=1.
\label{eqn:spectra;_normalisation}
\end{equation}

The effects of a region of charge noise can now be modelled using either approach (Eq.~\ref{eqn:multiple_fluctuators} or Eq.~\ref{eqn:multiple_fluctuators_integral_log}).  The choice will depend on exactly what information is known about the system.

In the following section we outline how to implement the superoperators and provide a discussion of some important limitations of this approach.

\section{Using the superoperators}
\label{sec:using_the_superoperators}
The superoperators in sections~\ref{sect:Determining_Q} and \ref{sect:multiple_fluctuators} were constructed on the basis of the commutation relation for our Heisenberg Hamiltonian (see Eq.~\ref{eqn:commutation_relation}).  The commutation relation meant we could express the total evolution operator as a product of two evolution operators, each describing the total time spent in one of the fluctuator states (see Eq.~\ref{eqn:evolution operator_for given_noise_history}).  The superoperators themselves also commute as a result.  Without further approximation the superoperators can be used to model individual processes satisfied by the Hamiltonian in Eq.~\ref{eqn:commutation_relation} (or a similar commuting Hamiltonian, such as the Ising interaction).

\begin{figure}[h]
\centerline{\includegraphics{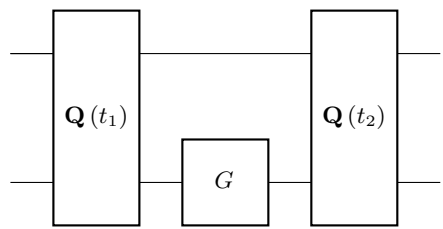}}
\caption{Multiple instances of the superoperator $\mathbf{Q}\left(t\right)$ separated by a non-commuting gate operation.  Attempting to use the superoperators to determine the effects of charge noise in a process like this can lead to the introduction of errors in the slow fluctuator regime.}
\label{fig:non_commuting_noisy_process}
\end{figure}

In more complex superoperator applications, for example when multiple applications of the superoperators are separated by non-commuting operations, it may be necessary to make a further approximation.  When modelling these more complex processes (see Fig.~\ref{fig:non_commuting_noisy_process} for an example where two superoperators are separated by a non-commuting gate operation, $G$) a problem arises with the formalism when considering the slow fluctuator limit.
Specifically, the superoperator terms which should describe no fluctuations at all, actually account for the possibility of a fluctuation occurring between superoperator applications.  We refer to these terms as \emph{cross terms}.  As the fluctuation rate increases the Poissonian weighting of these cross terms in the overall superoperator reduces, hence reducing the cross terms significance.  We now consider a simple example which illustrates how these cross terms manifest themselves, before showing how an approximate solution can be constructed for the slow fluctuator regime by removing the cross terms.

When the superoperators do not commute as in the example shown in Fig.~\ref{fig:non_commuting_noisy_process}, two or more applications of the superoperators leads to the introduction of unphysical cross terms in the slow fluctuator limit.  This becomes apparent when we consider the action of the superoperator describing no fluctuations $\mathbf{Q}^{\left(\rm nu\right)}_0\left(t\right)$, which may be expanded in terms of the superoperators, $\mathbf{Q}^\pm_0\left(t\right)$, each describing one of the two fluctuator states $\pm\xi$, in the no fluctuator limit
\begin{equation}
\mathbf{Q}^{\left(\rm nu\right)}_0\left(t\right)=\frac{1}{2}\left[\mathbf{Q}^+_0\left(t\right)+\mathbf{Q}^-_0\left(t\right)\right].
\label{eqn:no_fluctaution_components}
\end{equation}
The cross terms from the product of two (or more) of these superoperators which sandwich non-commuting operations results in the description of a single (or multiple) fluctuation(s).  For example, consider the system in Fig.~\ref{fig:non_commuting_noisy_process}.  The gate operation $G$ does not commute with the superoperators
\begin{equation}
\mathbf{Q}_{\rm total}=\mathbf{Q}\left(t_2\right)\mathbf{G}\mathbf{Q}\left(t_1\right),
\label{eqn:example_from_figure}
\end{equation}
where $\mathbf{G}$ is the superoperator representation of the gate operation $G$. Expanding out each of the fluctuator superoperators using
\begin{multline}
\mathbf{Q}\left(t\right)=\mathbf{Q}^{\left(\rm u\right)}\left(t\right)\Big[p_0\left(\lambda t\right)\mathbf{Q}^{\left(\rm nu\right)}_0\left(t\right)\\
+p_{>0}\left(\lambda t\right)\mathbf{Q}^{\left(\rm nu\right)}_{>0}\left(t\right)\Big],
\label{eqn:Q_breakdown}
\end{multline}
and Eq.~\ref{eqn:no_fluctaution_components}, with some rearranging we find
\begin{multline}
\mathbf{Q}_{\rm total}=\frac{p_0\left[\lambda \left(t_1+t_2\right)\right]}{4}\mathbf{Q}^{\left(u\right)}\left(t_2\right)\big[\mathbf{Q}^+_0\left(t_2\right)\mathbf{G}\mathbf{Q}^+_0\left(t_1\right)\\
+\mathbf{Q}^+_0\left(t_2\right)\mathbf{G}\mathbf{Q}^-_0\left(t_1\right)+\mathbf{Q}^-_0\left(t_2\right)\mathbf{G}\mathbf{Q}^+_0\left(t_1\right)\\
+\mathbf{Q}^-_0\left(t_2\right)\mathbf{G}\mathbf{Q}^-_0\left(t_1\right)\big]\mathbf{Q}^{\left(u\right)}\left(t_1\right)+\ldots~,
\label{eqn:no_fluctuations_cross_terms}
\end{multline}
where we have only shown the terms which should describe no fluctuations.  This entire expression should represent the total superoperator describing no fluctuations.  However, careful inspection shows the presence of two cross terms, which actually imply the occurrence of a fluctuation during the non-commuting gate operation.  Cross terms of this form are actually a manifestation of this superoperator formalism and should be removed without also removing any unitary evolution.

It should be emphasised that this problem only occurs in the slow fluctuator limit, where there is a significant probability of there being no fluctuations during a two-qubit operation.  As the fluctuation rate increases, the probability of a fluctuation occurring during the SINGLE qubit gate increases, which means that each application of the two-qubit gate becomes statistically independent.  In this limit, the formalism as presented so far is exact and does not require any attention to cross terms.

We now wish to remove cross terms describing processes which should not occur, such as those in Eq.~\ref{eqn:no_fluctuations_cross_terms}.  When two or more applications of the $\mathbf{Q}^{\left(\rm nu\right)}_0\left(t\right)$ superoperator occur in succession, the introduction of these unphysical terms also occurs.  It is possible to construct an approximate solution by carefully removing these cross terms.

Cross terms in the large $\lambda$ limit do not pose a problem as the $\mathbf{Q}^{\left(\rm nu\right)}_{>0}\left(t\right)$ superoperator provides the dominant contribution to $\mathbf{Q}_{\rm total}$ in this limit.  It will therefore be most important to remove the cross terms due solely to the $\mathbf{Q}^{\left(\rm nu\right)}_0\left(t\right)$ superoperator - the zeroth order cross terms.  In general, when many applications of the superoperators are required, there will be higher order cross terms.  For example, the first order cross terms would contain the $\mathbf{Q}^{\left(\rm nu\right)}_0\left(t\right)$ superoperator and a single instance of the $\mathbf{Q}^{\left(\rm nu\right)}_{>0}\left(t\right)$ superoperator.  Successive improvements to the approximate solution are achieved by removing these higher order cross terms.  As more orders of these terms are removed the approximation improves for larger $\lambda$, with the actual region of improvement dependent on the Poissonian weighting of the cross terms being removed.

As an example, the zeroth order cross terms, denoted $\mathbf{X}_0$, are removed by firstly re-weighting $\mathbf{Q}_{\rm total}$ using the Poisson distribution over the total process duration $T_{\rm total}$ (i.e., including non-commuting operations),
\begin{equation}
\mathbf{Q}_{\rm total}=p_0\left(\lambda T_{\rm total}\right)\mathbf{Q}_{\rm total}+p_{>0}\left(\lambda T_{\rm total}\right)\mathbf{Q}_{\rm total}.
\label{eqn:qtotal_reweighting}
\end{equation}
The cross terms can now be removed
\begin{multline}
\mathbf{Q}_{\rm total}\approx p_0\left(\lambda T_{\rm total}\right)\left(\mathbf{Q}_{\rm total}-\mathbf{X}_0\right)\\
+p_{>0}\left(\lambda T_{\rm total}\right)\mathbf{Q}_{\rm total},
\label{eqn:qtotal_cross_terms_removed}
\end{multline}
taking care to not also remove any unitary evolution.  The cross terms may also be removed from the second term, however the improvement from doing this is minimal due to the small contribution from the zeroth order terms at large $\lambda$.

\section{Conclusion}
The exchange interaction is of fundamental importance for controllable quantum mechanics in solid-state systems.  Its application to mediate spin flips or entanglement has particular importance in spintronics and QIP, hence the stability of the exchange interaction is crucial for precise time varying control.  In solid-state spin systems this stability can be affected by the local charge environment, in particular charge fluctuators, due to the exchange couplings dependence on the electron wave function overlap.

We have developed a model to describe the effect of charge fluctuators on the exchange interaction as a function of time, using superoperators dependent on the noise amplitude and rate.  These superoperators can be included in time-dependent calculations of the state of the spin system to model the effect of the charge noise.  Furthermore this analysis holds for other spin couplings, like the Ising interaction, where a commutation relation analogous to Eq.~\ref{eqn:commutation_relation} exists.

In the fast fluctuator limit we demonstrated how interaction with the environment via an exchange like decoherence channel leads to purely Markovian decoherence, although the decoherence operator leads to correlated noise across the two spins.

The generalisation to multiple fluctuators means that the effect of charge fluctuators distributed according to a spectral distribution function can also be modelled.  In the simpler case where only a small number of well defined fluctuators exist, the total superoperator is just the product of the individual fluctuator superoperators.

As our model is completely analytic, the effects of exchange fluctuations can in most instances be included trivially in more sophisticated analyses, without the need to explicitly sum over noise histories.  This is important for analysing the operation of spintronic devices as well as QEC and fault-tolerance for QIP.

\section*{Acknowledgements}
This work was supported by the Australian Research Council, the Australian Government and the US National Security Agency (NSA) and the Army Research Office (ARO) under contract number W911NF-08-1-0527.  JHC acknowledges the support of the Alexander von Humboldt Foundation and LCLH is the recipient of an Australian Research Council Australian Professorial Fellowship (DP0770715)


\end{document}